\documentclass[3p,times,twocolumn]{elsarticle}

\usepackage{ecrc}

\volume{00}

\firstpage{1}

\journalname{Nuclear Physics B Proceedings Supplement}

\runauth{S. Kawabata}

\jid{nuphbp}

\jnltitlelogo{Nuclear Physics B Proceedings Supplement}

\usepackage{amssymb}
\usepackage{amsthm}
\usepackage{multirow}

\usepackage[figuresright]{rotating}

\begin{document}

\begin{frontmatter}



\dochead{}

\title{New method for precise determination of top quark mass at LHC}

\author{Sayaka Kawabata}

\address{Department of Physics, Tohoku University, Sendai 980-8578, Japan}


\begin{abstract}
A new method to measure the mass of the top quark at the LHC is presented~\cite{Kawabataa:2014osa}. 
This method uses lepton energy distribution and ideally does not depend on the velocity distribution of the top quark.
We perform a simulation analysis of the top quark mass reconstruction using this method at the leading order, taking account of experimental circumstances.
We estimate the sensitivity of the mass determination.
The results show that this method is viable in realistic experimental conditions and has a possibility to achieve a good accuracy in determining a theoretically well-defined top quark mass by including higher-order corrections.
\end{abstract}

\begin{keyword}
top quark \sep top quark mass \sep measurement \sep LHC \sep QCD


\end{keyword}

\end{frontmatter}


\section{Introduction}
\label{sec:intro}
The mass of the top quark is an important input parameter to various physics.
In the electroweak precision tests, the top quark mass gives a large contribution to radiative corrections, and accordingly, its precise value is desired in order to scrutinize possible deviations from the Standard Model (SM)~\cite{Baak:2012kk,Baak:2013fwa}.
Furthermore, the stability of the SM vacuum up to the Planck scale depends crucially on the value of the top quark mass~\cite{Degrassi:2012ry,Buttazzo:2013uya}.
Now that the Higgs boson has been discovered~\cite{Aad:2012tfa,Chatrchyan:2012ufa} and the SM is getting more established, a demand for precise measurements of the top quark mass is increasing.

The top quark mass has been measured at the LHC and Tevatron, and their recent combined result yields~\cite{ATLAS:2014wva}
\begin{equation}
	m_t ~=~ 173.34 \pm 0.27\mbox{(stat)} \pm 0.71\mbox{(syst)} ~\mbox{GeV}\,.
\end{equation}
It achieves $0.4$\% precision, and more accurate results are expected to be obtained in future analyses~\cite{CMS:2013wfa}.

This mass, however, is not identical to the pole mass nor well-defined in perturbative QCD.
Since the above measurements utilize Monte Carlo (MC) simulations and reconstruct the mass from final-state momenta including jet momenta, the measured mass depends on the hadronization models in the MCs, which we cannot treat within perturbative QCD~\cite{Skands:2007zg}.
Thus, the definition of the measured mass in perturbative theory is ambiguous, and even its relation to the pole or $\overline{\rm MS}$ mass is not known~\cite{Hoang:2008xm}.

Some alternative methods have been proposed and developed to complement the above measurements with different systematic uncertainties or extract a theoretically well-defined top quark mass~\cite{Abazov:2011pta,Chatrchyan:2013haa,Aad:2014kva,ATLAS:topology,Chatrchyan:2013boa,Hill:2005zy,Aaltonen:2009hd,CMS:2013cea,Kharchilava:1999yj,Alioli:2013mxa,ATLAS:ttj}.
However, so far, no method has achieved to obtain a theoretically well-defined mass with high precision.

In this paper, we present a new method to measure the top quark mass at the LHC.
This method has characteristics of using lepton distributions and basically being independent of the kinematics of the production process of top quarks.
Consequently, it does not suffer from the ambiguity of hadronization models.
Using this method, we can determine the pole mass and $\overline{\mbox{MS}}$ mass of the top quark.

In Sec. \ref{sec:method}, a basis of our method, named the ``weight function method'', is presented.
We perform a simulation analysis of the top quark mass reconstruction using it and the results are shown in Sec.~\ref{sec:simulation}.
Section~\ref{sec:conclusions} is devoted to conclusions.

More details of the analysis and discussions in this paper are given in Ref.~\cite{Kawabataa:2014osa}.

\section{Weight function method}
\label{sec:method}
The weight function method is a method to measure various physical parameters in theory, proposed in Refs.~\cite{Kawabata:2011gz,Kawabata:2013fta}, and has a variety of applications besides a mass reconstruction of the top quark.

Consider the case that a parent particle $X$ decays into at least one lepton $\ell$ ($=e$ or $\mu$) plus any other particles ($X \rightarrow \ell + \mbox{anything}$). We suppose the parent particle $X$ is scalar or unpolarized (with respect to the direction of its boost) and the mass of the lepton can be neglected. Then, we can construct the following integral $I(m)$ using the (normalized) lepton energy distribution $D(E_\ell)$ in the laboratory frame:
\begin{equation}
	I(m)=\int dE_\ell D(E_\ell) W(E_\ell,m)\,,
	\label{eq:integral}
\end{equation}
where $W(E_\ell,m)$ is called a weight function and defined by
\begin{equation}
	\!\!\!\!\!\!\!\!W(E_\ell,m)=\!\int \!dE \left.\mathcal{D}_0(E;m)\frac{1}{EE_\ell} \,({\rm odd~fn.~of~}\rho)\right|_{e^{\rho}=E_\ell/E}
	\label{eq:weightfn}
\end{equation}
with the normalized lepton energy distribution $\mathcal{D}_0(E;m)$ in the rest frame of the parent particle. 
The $m$ is an arbitrary parameter included in $\mathcal{D}_0$ and supposed to be measured. 
We assume the theoretical expression of $\mathcal{D}_0$ can be obtained, and thus, weight functions are calculable theoretically.

As proved in Refs.~\cite{Kawabata:2011gz,Kawabata:2013fta}, the weighted integral $I(m)$ has the following property: $I(m)=0$ when the parameter $m$ takes its true value, that is, $I(m=m^{\rm true})=0$. The nontrivial point is that this property holds true independently of the Lorentz frame of the lepton energy distribution $D(E_\ell)$, and therefore, independent of velocities of the parent particle.
This means that only from lepton energy distribution in any frame, we can obtain the true value of $m$ as the zero of $I(m)$ without knowledge of the velocity distribution of the parent particle.
This has a great advantage in aiming at precision measurements at hadron colliders, where most processes of interest have missing particles and/or jets in their final states and theoretical predictions of production processes require parton distribution functions (PDFs) with comparatively large uncertainties, which make it difficult to reconstruct the kinematics of parton-level processes accurately.

In practice, lepton energy distribution in the laboratory frame is distorted by various experimental effects such as detector acceptance, event selection cuts and backgrounds.
In Sec.~\ref{sec:simulation}, we take these effects into account and develop a weight function method adapted to practical top mass reconstruction at the LHC.

\section{Simulation analysis of top mass reconstruction}
\label{sec:simulation}
\subsection{Weight function method for top mass reconstruction}
We apply the weight function method to the top quark mass reconstruction at the LHC.
The theoretical prediction of the longitudinal polarization of top quarks produced via $t\overline{t}$ at the LHC is $0.003$ induced only by the weak interaction within the SM~\cite{Bernreuther:2013aga}, and we can ignore it in a good approximation.
Thus, the top quark decaying leptonically at the LHC satisfies the requisites for the weight function method mentioned in the previous section.

\begin{figure}[t]
\includegraphics[width=6cm]{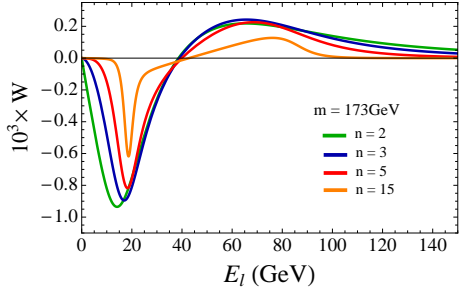}
\vspace*{-2mm}
\caption{\small
Weight functions $W(E_\ell,m)$ used in the analysis with $m=173$\,GeV, corresponding to $n=2,\,3,\,5$ and $15$ in Eq.~(\ref{eq:oddfunc}).
\label{fig:weightfn}
}
\vspace{-0mm}
\end{figure}

The theoretical prediction of the normalized lepton energy distribution in the rest frame of the top quark with a mass $m$, which we insert into Eq.~(\ref{eq:weightfn}) to construct weight functions, is given at the leading order (LO) by
\begin{eqnarray}
	\!\!\!\!\!\!\!\!\!\!\mathcal{D}_0(E;m)&\propto&E\left\{ \frac{m}{2}\left( 1-\frac{m_b^2}{m^2}\right)-E\right\}\nonumber\\
	&&\times\left\{\arctan \!\left(\frac{m_W}{\Gamma_W}\right)
	-\arctan\! \left(\frac{m_W^2-\mu^2_{\rm max}}{m_W \Gamma_W}\right)\right\}\nonumber\\
	&&\times\,\,\theta(\,0< E<E_{\rm max}\,)\,,
\end{eqnarray}
with
\begin{eqnarray}
	E_{\rm max}&\equiv&\frac{m^2-m_b^2}{2m}\,,\\
	\mu^2_{\rm max}&\equiv&\frac{2E(m^2-m_b^2-2mE)}{m-2E}\,,
\end{eqnarray}
where $m_b$, $m_W$, and $\Gamma_W$ represent the masses of the bottom quark, $W$ boson, and the width of the $W$ boson, respectively.
From this expression, we obtain weight functions $W(E_\ell,m)$.
In Fig.~\ref{fig:weightfn} we show the weight functions in the case we choose the odd function of $\rho$ in Eq.~(\ref{eq:weightfn}) as
\begin{equation}
	({\rm odd~fn.~of~}\rho)\,=\,n\tanh (n\rho)/\!\cosh(n\rho)\,,
	\label{eq:oddfunc}
\end{equation}
with $n=2,\,3,\,5$ and $15$.

By including higher-order radiative corrections in $\mathcal{D}_0$, we can obtain, in principle, the pole and $\overline{\mbox{MS}}$ masses with this method.
Since this method requires only lepton energy distribution and does not utilize jet momenta, it is free from the ambiguity of hadronization models, which enables a determination of a theoretically well-defined top quark mass.
In this first analysis, however, we concentrate on experimental aspects, which are a more difficult problem to tackle, and study at LO.
Considering higher-order corrections is delegated to our future work.

\subsection{Effects of lepton cuts and compensating method}
We perform a simulation analysis of the top mass reconstruction with the weight function method at the LHC to investigate how robust this method is in real experimental conditions.

We consider for signal events the production of $t\overline{t}$ which decay into a muon plus jets with $\sqrt{s}=14$\,TeV, and generate the signal and several major background events for the simulation analysis.
For the details of the event generation and event selection cuts in this analysis, see Ref.~\cite{Kawabataa:2014osa}.

Among many experimental effects which cause distortion to lepton energy distribution, the most serious effects are caused by the lepton cuts:
\begin{equation}
	p_T(\mu)>20{\rm \,GeV},~~|\eta(\mu)|<2.4\,,
	\label{eq:leptoncuts}
\end{equation}
where $p_T(\mu)$ and $\eta(\mu)$ are the transverse momentum and pseudo-rapidity of a muon, respectively.
The zero of $I(m)$ moves from the true value of the top quark mass significantly due to these lepton cuts.
To solve this problem, we pay attention to the following two points.
The independency of the parent particle velocities mentioned in Sec.~\ref{sec:method} is guaranteed only when the angular distribution of the lepton is flat in the rest frame of the parent particle (which is derived from our assumption that the parent particle is scalar or unpolarized).
Hence, we want to restore the flat distribution in order to take advantage of this method.
In addition, lepton distributions can be predicted with a relatively good accuracy.
Considering these points, we choose to compensate for the loss caused by the lepton cuts, using MC simulation events.

\begin{figure}[t]
\includegraphics[width=6cm]{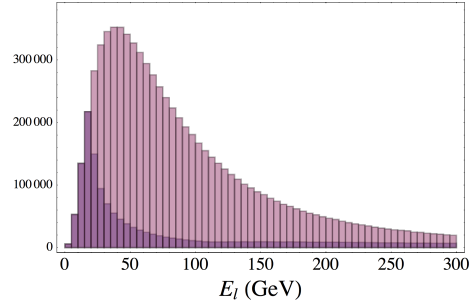}
\vspace*{-2mm}
\caption{
Sum of the lepton energy distributions of the compensated events (dark purple) and the events after all the cuts (light pink).
\label{fig:compensate}
}
\vspace{-0mm}
\end{figure}

Figure~\ref{fig:compensate} illustrates the compensating procedure, showing the sum of the lepton energy distributions of the compensated events and the signal events after all the cuts.
The compensated events satisfy the complement of the above lepton cuts, namely $p_T(\mu)<20{\rm \,GeV}$ or $|\eta(\mu)|>2.4$\,.
To fix the normalization of the compensated events, we do not want to use the whole distribution or cross section of events, which are accompanied with inevitable uncertainties from PDFs.
For this reason, we determine the normalization so that the $p_T(\mu)$ distributions are connected smoothly around $p_T(\mu)=20$\,GeV.

Note that we have to assume some value for the top quark mass of the compensated MC events.
We call it $m_t^c$.
The reconstructed mass is obtained from the following consistency condition: if $m_t^c$ coincides with the input mass, (which corresponds to the true mass of data in the case of real experiments,) the zero of $I(m)$ should be $m_t^c$, namely, $I(m=m_t^c)=0$ for $m_t^c=m_t^{\rm input}$.

\begin{figure}[t]
\includegraphics[width=6cm]{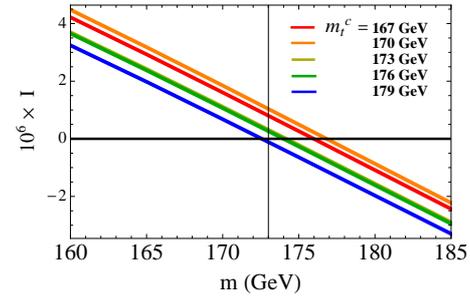}
\vspace*{-2mm}
\caption{
Weighted integrals $I(m)$ with various $m_t^{\rm c}$ after all the cuts. The weight function used corresponds to $n=2$ in eq.~(\ref{eq:oddfunc}). The input value of the top quark mass is $173$\,GeV.
\label{fig:ImWithmtcAftCuts}
}
\vspace{-0mm}
\end{figure}

Figure~\ref{fig:ImWithmtcAftCuts} shows $I(m)$ with various $m_t^c$.
The $m_t^{\rm input}$ is $173$\,GeV for this figure.
Although the top mass of the compensated events $m_t^c$ vary from $167$ to $179$\,GeV, the zeros of $I(m)$ vary less.
Thus, we can see that $I(m)$ does not depend strongly on the top mass of the compensated events.
This helps us to extract the true mass.

\begin{figure}[t]
\includegraphics[width=6cm]{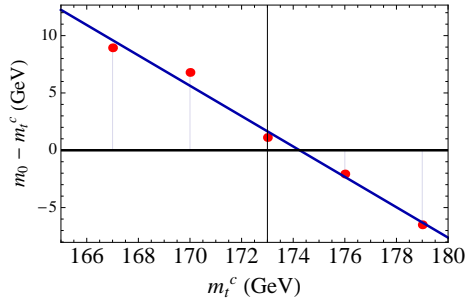}
\vspace*{-2mm}
\caption{
The zero of $I(m)$, $m_0$, minus $m_t^{\rm c}$ as a function of $m_t^{\rm c}$ (red points). The weight function used corresponds to $n=2$. The input value of the top quark mass is $173$\,GeV. The blue line shows the linear function fitted to the red data points.
\label{fig:m0-mtcAftCuts}
}
\vspace{-0mm}
\end{figure}

Figure~\ref{fig:m0-mtcAftCuts} shows the zero of $I(m)$, $m_0$, minus $m_t^c$ as a function of $m_t^c$.
According to the above consistency condition, the reconstructed mass is at the value of $m_t^c$ where $m_0$ coincides with $m_t^c$.
Therefore, from Fig.~\ref{fig:m0-mtcAftCuts}, we obtain the reconstructed mass as $174.2$\,GeV, the zero of the blue line for this simulation analysis.
The deviation from the input mass, $1.2$\,GeV, is understood as a MC statistical error due to a limited number of generated events, and effects of the top quark width which this method does not take into account.
Thus, this method works within these errors, even after including the effects of event selection cuts.

\subsection{Sensitivity of mass determination}
\begin{figure}[t]
\includegraphics[width=6cm]{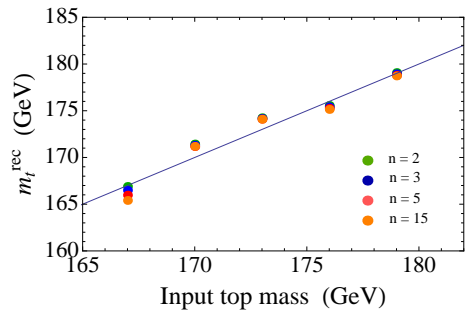}
\vspace*{-2mm}
\caption{
Reconstructed top quark mass as a function of the input mass. The weight functions used correspond to $n=2,\,3,\,5$ and $15$. The blue line shows the line where the reconstructed mass is equal to the input mass.
\label{fig:MtMeasAftCuts}
}
\vspace{-0mm}
\end{figure}

Our strategy is to subtract background distributions (accompanied by statistical and systematic errors), compensate for the loss caused by the lepton cuts, and reconstruct the top quark mass according to the procedure described in the previous section.
We perform the same simulation analysis of the mass reconstruction for the various input top quark masses as in the previous section.
The results are shown in Fig.~\ref{fig:MtMeasAftCuts} for several choices of weight functions.
Each reconstructed top mass agrees well with the corresponding input mass, and its deviation is consistent with the effects of the MC statistical error and top quark width.

We evaluate the sensitivity of the mass determination.
We consider signal statistical errors and several major sources of systematic uncertainties.
Table~\ref{tab:uncertainties} summarizes the estimates of uncertainties.
The estimated statistical errors are about $0.4$\,GeV with an integrated luminosity of $100$\,fb$^{-1}$ for the sum of the lepton($e,\mu$)$+$jets events.
The uncertainties involved with jet energy scale (JES) contribute through event selection cuts concerning jets and are fairly small reflecting the characteristics of this method using only lepton distributions.
The uncertainties from factorization scale dependence in PDF and hadronization processes arise from the compensating method.
Although the factorization scale uncertainties are dominant with the size of $1.5$\,GeV, this analysis is at LO and these uncertainties are expected to be reduced by including the NLO corrections.

\begin{table}[t]
	\centering
	\begin{tabular}{cc|cccc}
		\hline 
		\multicolumn{2}{c}{}&~Signal~&Fac. scale&~~JES~~&$\!\!\!\!$Background\\
		\multicolumn{2}{c}{}&stat. error&(signal)&(signal)&stat. error\\
		\hline 
		\multirow{4}{*}{$\!\!n\!\!\!$} & 2 & $0.4$& $+1.5/\!-\!1.6$ & $+0.0/\!-\!0.1$ & $0.4$\\ 
		& 3 & $0.4$& $+1.5/\!-\!1.5$ & $+0.1/\!-\!0.3$ & $0.4$\\ 
		& 5 & $0.5$ & $+1.4/\!-\!1.4$ & $+0.2/\!-\!0.4$ & $0.5$\\
		& 15 & $0.5$ & $+1.5/\!-\!1.3$ & $+0.2/\!-\!0.6$ & $0.6$\\ 
		\hline
	\end{tabular}
	\caption{\label{tab:uncertainties} Estimates of uncertainties in GeV from several sources in the top mass reconstruction. The weight functions used correspond to $n=2,\,3,\,5$ and $15$. The input value of the top quark mass used in the estimates is $173$\,GeV. The signal statistical errors correspond to those with an integrated luminosity of $100$\,fb$^{-1}$ and for the sum of the lepton($e,\mu$)+jets events. The background statistical errors are also for $100$\,fb$^{-1}$.}
\end{table}

The most important subject we should deal with next is to include theoretical corrections in this analysis.
Including the NLO and NNLO corrections enables us to extract the pole and $\overline{\rm MS}$ mass of the top quark with this method.
Note that owing to the boost-invariant nature of this method, important corrections are basically to the decay process of the top quark.
Since this method assume the on-shell top quark, the effects of the off-shellness also should be incorporated.

\section{Conclusions}
\label{sec:conclusions}
A new method to measure the top quark mass at the LHC is presented.
This method is based on the weight function method proposed in Refs.~\cite{Kawabata:2011gz,Kawabata:2013fta} and uses lepton energy distribution.
The experimental viability is investigated at LO and the sensitivity of the mass determination is estimated.
The estimated statistical error is about $0.4$\,GeV with an integrated luminosity of $100$\,fb$^{-1}$ at $\sqrt{s}=14$\,TeV.
We also estimate some of the major systematic uncertainties and find that they are under good control. 
By including higher-order corrections in this method, the pole and $\overline{\rm MS}$ mass of the top quark can be determined.

\section*{Acknowledgements}
This analysis is performed in collaboration with Y. Shimizu, Y. Sumino and H. Yokoya, and I would like to thank all of them.
This work is supported by Grant-in-Aid for JSPS Fellows under the program number 24$\cdot$3439.




\nocite{*}
\bibliographystyle{elsarticle-num}
\bibliography{martin}


\end{document}